\documentstyle[psfig,graphics]{mn}

\def\simlt{\lower.5ex\hbox{$\; \buildrel < \over \sim \;$}}
\def\simgt{\lower.5ex\hbox{$\; \buildrel > \over \sim \;$}}
\def\l#1{\left#1}
\def\eg{{\rm e.g.$\,$}}
\def\ie{{\rm i.e.$\,$}}
\def\cf{{\rm cf.$\,$}}
\def\etal{{\rm et al.$\,$}}
\def\r#1{\right#1}

\def\apj{{\rm ApJ}}
\def\apjs{{\rm ApJS}}

\def\aa{{\rm A\&A}}
\def\mnras{{\rm MNRAS}}

\def\wtilde{\widetilde}

\begin{document}

\title[Gravitational lensing of CMB anisotropies]{Gravitational lensing of cosmic microwave background
anisotropies and cosmological parameter estimation}

\author[R. Stompor, G. Efstathiou]{R. Stompor{\thanks{on the leave of
absence from Copernicus Astronomical Centre, Warszawa, Poland}} and G. Efstathiou\\
Institute of Astronomy, Madingley Road, Cambridge CB3 OHA\\
E-mail: radek@ast.cam.ac.uk; gpe@ast.cam.ac.uk
}

\date{\today}

\maketitle

\begin{abstract}
Gravitational lensing, caused by matter perturbations along the
line-of-sight to the last scattering surface, can modify the shape of
the cosmic microwave background (CMB) anisotropy power spectrum. We
discuss the detectability of lensing distortions to the
temperature, polarisation and temperature-polarisation
cross-correlation power spectra and we analyse
how lensing might affect the estimation of cosmological parameters.
For cold dark matter-like models with present-day matter power spectra
normalised to match the abundances of rich clusters of galaxies, 
gravitational lensing causes detectable distortions
to  cosmic variance limited CMB experiments sampling high multipoles ($\ell
\simgt 1000$).  Gravitational lensing of the CMB, although
a small effect, allows independent determinations of
the curvature of the universe  and the cosmological constant,
\ie  breaking the  so-called {\it geometrical degeneracy} in CMB
parameter estimation discussed by Bond, Efstathiou \&\ Tegmark (1997)
and Zaldarriaga, Spergel \&\ Seljak (1997). Gravitational lensing of
the CMB temperature and polarisation patterns should be detectable by 
the Planck Surveyor satellite leading to useful independent constraints
on the cosmological constant and spatial curvature.

\end{abstract}

\begin{keywords}
cosmic microwave background anisotropies -- gravitational lensing, 
cosmological parameters estimation
\end{keywords}

\setcounter{figure}{0}
\setcounter{table}{0}

\section{Introduction and Motivation}

Since the early papers on the cosmic microwave background anisotropies 
(CMB) by   Peebles \&\ Yu (1968), Do\-ro\-shke\-vich,
Zel'dovich, \&\ Sunyaev (1978),  Wilson \&\ Silk (1981) and others, 
it has been evident that the  CMB
anisotropies are  sensitive to 
fundamental cosmological parameters. These include parameters
that define the background cosmology 
(such as the geometry and matter content) and parameters
that define the nature of irregularities in the early Universe
(such as the amplitude and shape of the fluctuation spectrum).
Early attempts to constrain the parameters of cold dark matter (CDM)
models were made by  Bond \&\ Efstathiou (1984) and
Vittorio \&\ Silk (1984). More recently, the parameters of 
CDM-type models have been constrained using the 
$COBE$--DMR data alone (e.g. Bennett \etal 1996, Bunn, Scott \&\ White 1995, Stompor, G\'orski \&\ Banday
1995, G\'orski \etal 1998), and $COBE$ combined with degree-scale measurements of CMB anisotropies
(e.g. Hancock \etal 1997, Lineweaver \etal 1997, Bond \& Jaffe 
1997). 

In the near future, long-duration balloon flights and satellite
experiments promise to provide a wealth of high quality data on the
CMB anisotropies. This has stimulated a number of theoretical
investigations on the determination of cosmological parameters from
observations of the CMB anisotropies (\eg Jungman \etal 1996,
Bersanelli \etal 1996, Bond, \etal 1997,
Zaldarriaga \etal 1997, Efstathiou \&\ Bond 1998, Eisenstein, Hu \&\
Tegmark 1998).
These studies have confirmed that many cosmological parameters, or
combinations of parameters, can be determined by future satellite
missions to unprecedented precisions of a few percent or
better. However, these studies have identified some degeneracies
between sets of cosmological parameters\footnote {\ie parameter
sets that lead to almost indistinguishable CMB power spectra.}
estimated from the linear CMB power spectra alone.  Since the entire
statistical information on the CMB anisotropies in Gaussian theories
is contained in the power spectrum, such parameter degeneracies impose
serious limitations on the ability of CMB experiments to constrain
cosmological parameters without invoking additional external constraints.

In particular, Bond \etal (1997) and Zaldarriaga \etal (1997) have
emphasized that cosmological models with identical fluctuation
spectra, matter content and angular diameter distance to the
scattering surface (see Section 2.1 below) will produce statistically
almost indistinguishable power spectra of CMB fluctuations.  This property
(which we call the {\it geometrical degeneracy} hereafter) means that
in the limit of validity of linear perturbation theory, CMB
measurements cannot set strong independent bounds on the spatial
curvature and cosmological constant and hence cannot unambiguously
constrain the spatial geometry of the Universe.

In fact there are many additional observational constraints that can
be used to break the geometrical degeneracy. Examples include accurate
measurements of the Hubble constant, the age of the Universe and the
geometrical constraints imposed by Type Ia supernovae light curves
[see Figure \ref{fig1} and the more detailed discussions by White
(1998), Tegmark, Eisenstein \&\ Hu (1998) and Efstathiou \&\ Bond (1998)].  
However, before invoking  more conventional astronomical observations, it
is worthwhile analysing whether there are non-linear contributions to
the CMB anisotropies that can break the geometrical degeneracy. If
such effects are present, then it may be possible to break the geometrical
degeneracy using measurements of the CMB alone. In this
paper, we analyse the effect of gravitational lensing on the CMB
anisotropies.  Although acknowledged to be small (Blanchard \&\ Schneider 1987,
Cole \&\ Efstathiou 1989, Sasaki 1989, Seljak 1996), the gravitational lensing effect may
be detectable by the high precision observations of the CMB
anisotropies expected from future satellite experiments.
The possibility of utilising
gravitational lensing to break the geometrical degeneracy has been
noticed independently by Metcalf \&\ Silk (1998). In this paper, we
analyse the effects of gravitational lensing on the temperature,
polarisation and temperature-polarisation cross-correlation 
power spectra and assess whether it is possible to
observe these effects with the  MAP (Bennett \etal 1997) and Planck
(Bersanelli \etal 1996) satellites.

\section{The geometrical degeneracy}

\subsection{Physical mechanism}

In this paper we restrict ourselves to cold dark matter (CDM)
cosmologies with adiabatic scalar perturbations, an arbitrary value of
the curvature ($\Omega_K\equiv -K/H_0^2$) and cosmological constant
($\Omega_\Lambda\equiv \Lambda/3 H_0^2$).  Following Bond \etal (1997)
we use physical densities, $\omega_i\equiv\Omega_ih^2$, to define the
matter content of the universe \footnote{Where h is the Hubble's
constant $H_0$ in units of $100\; {\rm km}{\rm s}^{-1}{\rm Mpc}^{-1}$.}, with
$i=K,\Lambda$,b,c,$\gamma$,$\dots$, and $\Omega_{\rm b}$, $\Omega_{\rm c}$,
$\Omega_{\gamma}$,$\dots$ are the density parameters of baryons,
cold dark matter (CDM), photons etc.
We assume the standard thermal history throughout this paper with
recombination at redshift $z\sim 1100$ (Peebles 1968) and ignore the
possibility of reionization. In the numerical examples described below,
we have assumed a scale-invariant (Harrison-Zel'dovich) power spectrum
of primordial scalar and adiabatic fluctuations \ie $\propto
\sqrt{k^2+K},$ where a wavenumber $k$ is the separation constant of
the Helmholtz equation (e.g. Harrisson 1967).

As is well known, the power spectrum of the CMB anisotropies in such
models displays prominent `Doppler' peaks (see, \eg Figures
2a,b). The shape of the CMB power spectrum and, in particular, the
locations and relative heights of the peaks, depend sensitively on
cosmological parameters (e.g. Hu \&\ Sugiyama 1995). The Doppler peak
structure is imprinted into the present day CMB power spectrum at the
time of recombination.  Since recombination occurs at a high redshift,
no plausible value of the cosmological constant or spatial curvature
can influences the dynamical evolution of the universe at that time
\footnote{We do not consider the possibility of a dynamically evolving
cosmological constant or ``quintessence''-like component as 
discussed  by Ratra \&\ Peebles (1988), Turner
\&\ White (1997), Caldwell, Dave \&\ Steinhardt (1998), Huey \etal (1998)}.  
The statistical
properties of the CMB anisotropies at the time of last scattering are
therefore determined by the form of the initial fluctuations spectrum
(mode, shape and amplitude) and by the physical densities that
determine the sound speed prior to recombination.

After last scattering, (assuming that the universe remains neutral)
the only mechanism that can
affect a freely-falling CMB photon is the gravitational interaction
with the evolving matter field. In the linear approximation 
this is sometimes called the integrated
Sachs-Wolfe effect [see, e.g. the review by Bond (1996)] and
is of importance only for temperature fluctuations on  the largest angular 
scales. Since the large-scale anisotropies have large statistical 
uncertainties (cosmic variance), the integrated Sachs-Wolfe effect
cannot break the geometrical degeneracy except for 
extreme values of the cosmological parameters [see Efstathiou \&\ Bond (1998)
for detailed calculations]. In linear
theory, the geometrical degeneracy can be assumed to be exact
for most practical purposes.

Two models will have statistically indistinguishable 
temperature, polarisation and cross-correlation linear power spectra
as a result of the geometrical degeneracy 
if they have:

\begin{description}

\item{(i)}{ identical matter content of those components
that determine the sound speed at recombination,
$\omega_{\rm c}$, $\omega_{\rm b}$, ...;}

\item{(ii)}{ identical comoving distance to
the last scattering surface ($r_{LS}$), where
\begin{eqnarray}
{r_{LS}\over 3000\;{\rm Mpc}}=
 {1\over \sqrt{\omega_K}}
\sinh\l[\int_{a_{LS}}^{a_0}{{\omega_K}^{1/2} da\over
\sqrt{\omega_Ka^2+\omega_\Lambda a^4+\omega_{\rm m}a}}\r],
\label{rlss}
\end{eqnarray}
$\omega_{\rm m}\equiv\omega_{\rm c}+\omega_{\rm b},$
(Bond \etal 1997). In equation (\ref{rlss}) we 
have assumed an open universe ($\omega_K > 0$).
The upper limit in (\ref{rlss}) is the value of the scale 
factor $a$ at the present epoch and the lower limit is
the value at the time of the last scattering [for which we use
the fitting formula of Hu \&\ White (1997)];}

\item{(iii)}{ identical fluctuation spectra normalised to have the 
same  amplitudes at the time of recombination.}
\end{description}

If the above conditions are satisfied, the CMB power spectra of the
two models will be indistinguishable on small angular scales, but
will differ at large angles because of geometrical effects on
near curvature scales and, in the  case of temperature anisotropies,
through the integrated Sachs-Wolfe effect. However, these effects are
weak discriminators of models.  

In the analysis that follows, when we compare two models,
we normalise them so that the 
root mean square mass deviation computed within a top
hat window of radius $R=8h^{-1}$Mpc 
[denoted hereafter $\sigma_R(t)$] is the same {\it at the time
of last scattering} ($t=t_{LS}$). This prescription determines the relative
normalisations of any two models that we wish to compare, but
does not determine the absolute normalisation.

The effects of gravitational lensing depend, of course, on the
absolute normalisation of the matter fluctuations. We therefore
normalise a given target model so that the rms mass fluctuations
within a sphere of $8h^{-1}$Mpc
at the present day ($t=t_0$) reproduces the abundances of rich clusters
of galaxies. We therefore impose the constraint,
\begin{equation}
\sigma_8 (t_0) =  \l(0.52\pm 0.12\r) \Omega_{\rm m}^{-0.47},
\label{cluster}
\end{equation}
from the recent analysis of Eke, Cole \&\ Frenk (1996).  The
normalisation of scale-invariant models derived from
(\ref{cluster}) is usually lower than that
inferred from the 4 year COBE-DMR  data (e.g. G\'orski \etal 1998).
However, equation (\ref{cluster}) provides a more direct measure of the 
amplitude of the mass fluctuations which generate
gravitational lensing effects at recent epochs.

We give the $1\sigma$ uncertainty in equation (\ref{cluster}), though this is
sufficiently small that it has no significant effect on results
described below. Some authors, e.g. Viana \&\ Liddle (1996)
deduce slightly larger values of $\sigma_8(t_0)$ for low density
models, in which case our analysis will underestimate the effects of
gravitational lensing on the CMB.

In summary, we pick a specific set of cosmological parameters to
define a target model and we use equation (\ref{cluster}) to set the absolute
normalisation of the fluctuation spectrum.  When we compare models
with a different set of cosmological parameters, we choose a
normalisation so that the fluctuation spectra have the same amplitude
at the time of recombination, so preserving the geometrical
degeneracy. 

\begin{figure}
\resizebox{\columnwidth}{!}{\includegraphics{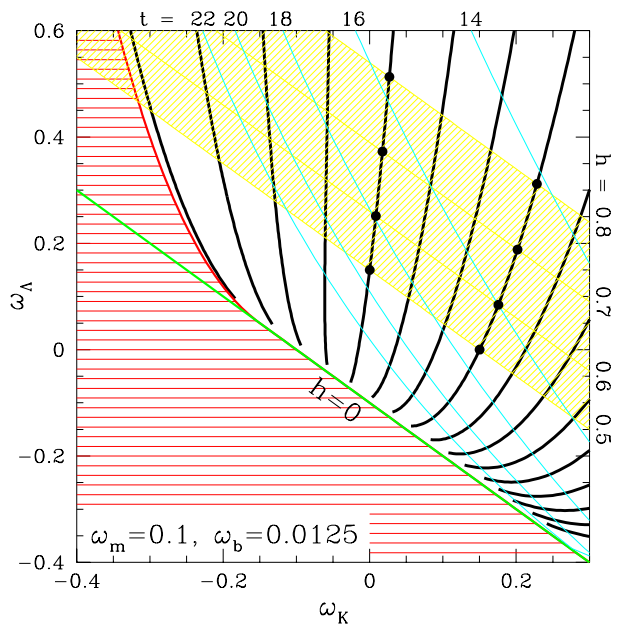}}
\caption{
The $\omega_K-\omega_\Lambda$ plane 
showing  the sets of models (curved thick solid lines) satisfying the
geometrical degeneracy and hence having nearly 
indistinguishable  power spectra according to 
linear perturbation theory.
These curves are computed for the models with $\omega_{\rm m}=0.1,$ and
$\omega_{\rm b}=0.0125$. Lines of constant Hubble parameter values (solid
thin straight lines labelled with Hubble parameter values at the right
edge of the figure), and of constant age (solid thin lines labelled at
the top of the figure with the age in Gyr) are also shown.
Filled circles show the two families of nearly degenerate
models with power spectra plotted in Figures 2a,b.
The parameters of these models are listed in Table 1.
}
\label{fig1}
\end{figure}

\begin{figure*}
\centerline{
\psfig{file=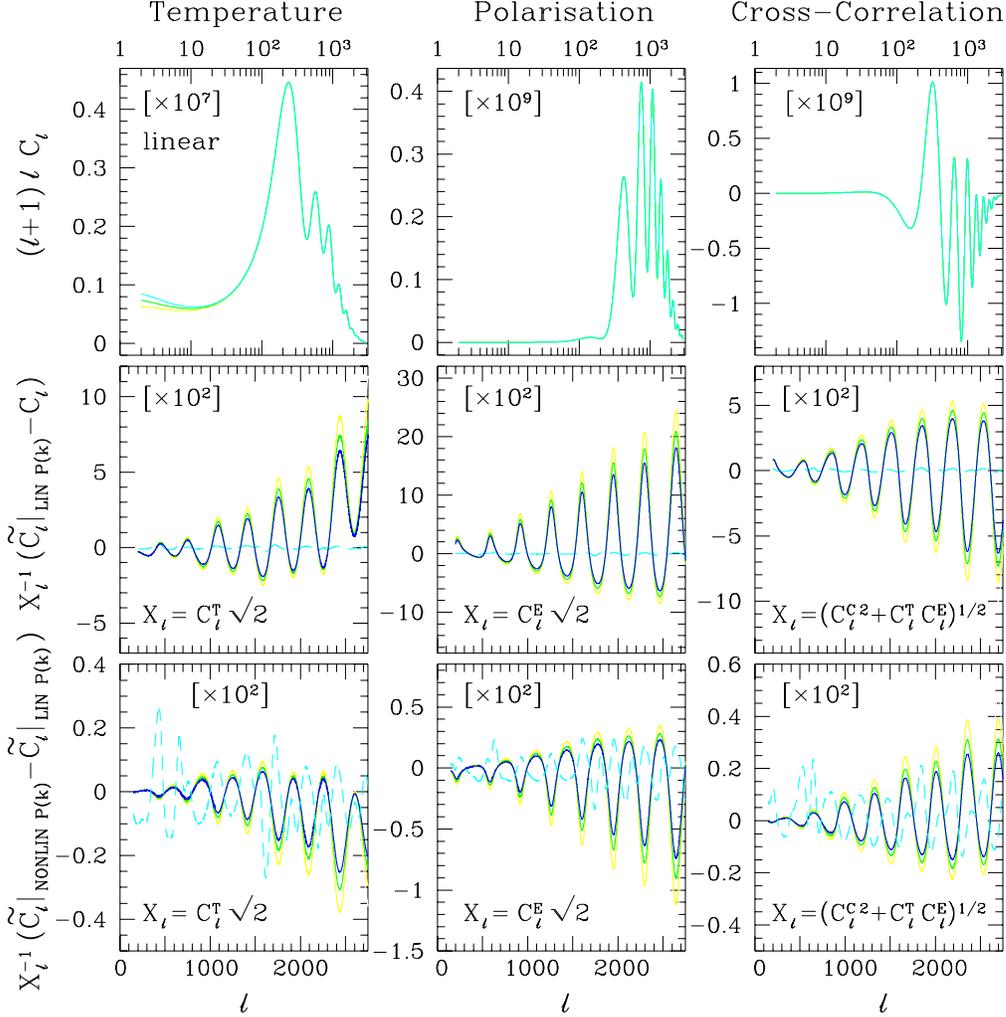,width=14cm,rheight=14cm,rwidth=14cm}
}
\caption{ 
{\bf(a)} The four lines in each panel of the {\bf top row} show nearly
identical linear temperature, polarisation and cross-correlation power
spectra for each of four nearly degenerate models
(models of class 1 listed in Table 1).
{\bf Middle row:} The solid lines show the gravitational lensing contribution
assuming only the linear matter power spectrum at all epochs.
The solid lines in the {\bf bottom panels} show the changes of
gravitational lensing contribution caused by the non-linear corrections
to the matter power spectrum evolution (implemented through the Peacock \&\
Dodds (1996) formalism).
The dashed lines in the middle and lower row show the typical differences
between linear radiation power spectra plotted in the top row.
Note that the $X_{\ell}$ quantity used to normalise the
gravitational lensing contributions in the two bottom rows is proportional
to the cosmic variance (\eg eqn.\,\ref{cosvar}).
The normalisation for target model (model 1a) is fixed according to
the present-day cluster abundances (eqn.$\,$2). The remaining models are required to
have, the same amplitude of the fluctuations at the time of last
scattering. The presented results have been multiplied by factors as given
in each panel.
}
\label{fig2a}
\end{figure*}

These points are illustrated in Figures \ref{fig1} and \ref{fig2a}.  Figure
\ref{fig1} shows degenerate loci in the $\omega_K$, $\omega_\Lambda$
plane ($r_{LS}= {\rm constant}$) for models with $\omega_{\rm m}=0.1$
and $\omega_{\rm b} = 0.0125$. Figures 2a,b show the linear CMB power
spectra for two sets of models
satisfying the geometrical degeneracy plotted as filled circles in
Figure \ref{fig1}. These have been computed using a version of the
CMBFAST code developed by Seljak \&\ Zaldarriaga (1996) which we have
modified to gain an improvement in  accuracy (see Section 4.2).  Clearly, the
spectra for each set of models are almost indistinguishable. The only
significant deviations are at low multipoles ($\ell\simlt 100$) and
are a consequence of the integrated Sachs-Wolfe effect described
above.  In fact, the numerically computed spectra in Figures 2a,b also show
some residual differences at high $\ell$ (illustrated by the dashed
line in the middle panels of Figure 2a,b).  However, these differences
are dominated by residual numerical inaccuracies in CMBFAST (see the discussion
in Section 4.2).  For models with reasonable normalisations 
[\ie reproducing cluster abundances as inferred from equation (\ref{cluster})]
these numerical errors are much smaller than differences arising from
gravitational lensing. 

\subsection{Degeneracy lines}

If we  keep all of the parameters of a target model
fixed but vary $\omega_K$ and 
$\omega_\Lambda$, the geometrical degeneracy will be satisfied if
\begin{equation}
\delta \omega_K \l({\partial r_{LS}\over \partial
\omega_K}\r)_{0} + \delta \omega_\Lambda\l({\partial r_{LS}\over \partial \omega_\Lambda}\r)_{0}
\simeq 0,
\label{degeqn}
\end{equation}
where the subscript $0$ on any quantity denotes that it is computed 
assuming the parameters of the target model. We can define two 
new parameters $\omega_\parallel$ and $\omega_\perp$, 
\begin{equation}
\l[
\begin{array}{c}
\displaystyle{\omega_{\parallel}}
\\
\displaystyle{\omega_{\perp}}
\end{array}
\r]\equiv
\l[
\begin{array}{c c}
\displaystyle{\cos\phi}  & \displaystyle{\sin\phi}\\
\displaystyle{-\sin\phi} & \displaystyle{\cos\phi}
\end{array}
\r]
\l[
\begin{array}{c}
\displaystyle{\delta\omega_K}
\\
\displaystyle{\delta\omega_\Lambda}
\end{array}
\r]
,
\label{newpars}
\end{equation}
where $\displaystyle{\phi\equiv -\arctan\l[\l({\partial r_{LS}\over
\partial \omega_\Lambda}\r)_{0}^{-1}\l({\partial r_{LS}\over \partial
\omega_K}\r)_{0}\r]}$ is the angle between a degenerate curve plotted
in Figure \ref{fig1} and the $\omega_K$ axis. Models satisfying
$\omega_\perp=0$ thus have the same value of $r_{LS}$ for small
variations of parameters and therefore satisfy the geometrical
degeneracy.  Hereafter we call any direction with $\omega_\perp=0$ a
degeneracy direction. If the geometrical degeneracy were perfect, the
derivative of the CMB power spectrum along a degeneracy direction
should be exactly equal to zero.  The numerical derivatives of linear
power spectra are discussed in section 4.2 (see also Efstathiou \&\ Bond 1998)
and shown in Figures 4 \&\ 8.

\setcounter{figure}{1}

\begin{figure*}
\centerline{
\psfig{file=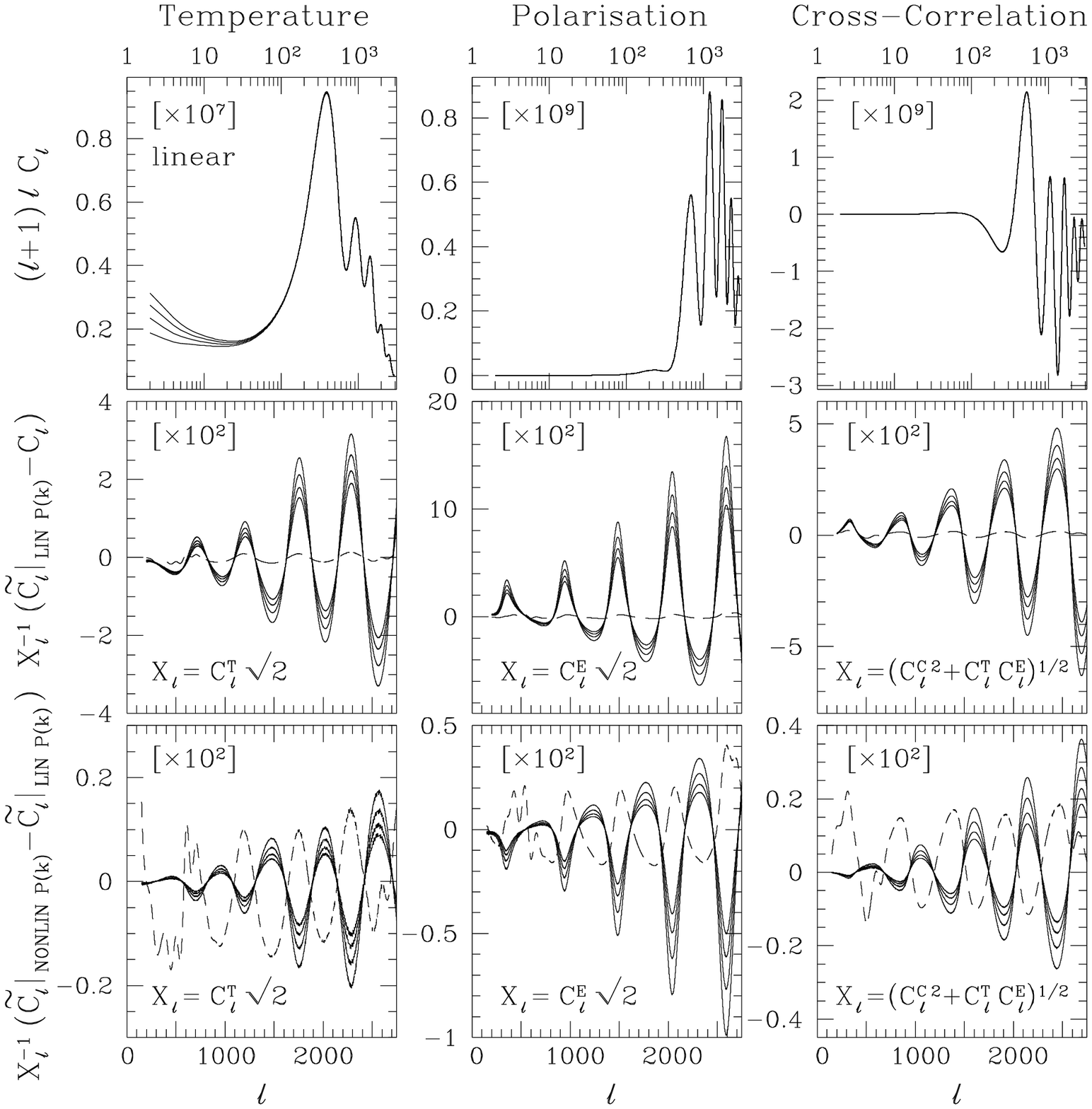,width=14cm,rheight=14cm,rwidth=14cm}
}
\caption{{\bf (b)} As figure 2a but for the second class of models
with parameters listed in Table 1.}
\label{fig2b}
\end{figure*}

\section{Gravitational lensing contribution}

\subsection{Formalism}

The gravitational lensing contribution to the temperature power
spectra has been computed in the past by several authors (e.g. 
Blanchard \&\ Schneider 1987, Cole \&\ Efstathiou 1989,
Sasaki 1989, Seljak 1996, Mart\'{\i}nez-Gonz\'{a}lez, Sanz \&\ Cay\'on
1997, Zaldarriaga \&\ Seljak 1998).
Here we follow the approach of Seljak (1996) and Zaldarriaga \&\
Seljak (1998).  We denote the linear power spectra as $C_\ell^T,$
$C_\ell^E$ and $C_\ell^C$ for temperature, polarisation (E-component
only) and their cross-correlation respectively. 
The CMB spectra including the  gravitational
lensing contribution are assigned a tilde. 

The full radiation power spectra including gravitational lensing can
be expressed as convolutions of the corresponding linear spectra
$(I=T,E,C)$,
\begin{equation}
\wtilde{C}^I_\ell=C^I_{\ell}+\sum_{\ell'}{\cal W}^I_{\ell\ell'}C^I_{\ell'}
\label{GLcl}
\end{equation}
where the window functions ${\cal W}^I_{\ell\ell'}$ are given as,
\begin{eqnarray}
{\cal W}^T_{\ell\ell'}&=&
{\ell'^3\over 2}\int_0^{\infty}d\,\theta\,\theta\,
\l[\sigma_2^2\l(\theta\r)J_2\l(\ell'\,\theta\r)\r.\nonumber\\
&&\l.-\sigma^2\l(\theta\r)J_0\l(\ell'\,\theta\r)\r]J_0\l(\ell\,\theta\r)\label{wT}\\
{\cal W}^E_{\ell\ell'}&=&{\ell'^3\over 4}\int_0^{\infty}d\,\theta\,\theta\,
\l\{ {\sigma_2^2\over2}\l(\theta\r)\l[J_2\l(\ell'\,\theta\r)+J_6\l(\ell'\,\theta\r)\r]\r.\nonumber\\
&&\l.-\sigma^2\l(\theta\r) J_4\l(\ell'\,\theta\r)\r\}J_4\l(\ell\,\theta\r)+{1\over 2}{\cal W}^T_{\ell\ell'}\label{wE}\\
{\cal W}^C_{\ell\ell'}&=& {\ell^3\over 2}\int_0^{\infty}d\,\theta\,\theta\,
\l\{ {\sigma_2^2\over2}\l(\theta\r)\l[J_0\l(\ell'\,\theta\r)+J_4\l(\ell'\,\theta\r)\r]\r.\nonumber\\
&&\l.-\sigma^2\l(\theta\r) J_2\l(\ell'\,\theta\r)\r\}J_2\l(\ell\,\theta\r)\label{wTE}
\end{eqnarray}
and photon path dispersions,
\begin{eqnarray}
\sigma^2\l(\theta\r)&=&16\pi^2\int_0^\infty dk k^3 \int_0^{\chi_{LS}}
\,d\chi\,P_{\phi}\l(k,\tau_0-\chi\r) \times\label{sigmath}\\
&\times&W^2\l(\chi,\chi_{LS}\r)\l[1-J_0\l(k\,\theta\sin_K\chi\r)\r];
\nonumber\\
\sigma_2^2\l(\theta\r)&=&16\pi^2\int_0^\infty\,dk\,k^3 \int_0^{\chi_{LS}}
\,d\chi\,P_{\phi}\l(k,\tau_0-\chi\r) \times \label{sigmath2}\\
&\times&W^2\l(\chi,\chi_{LS}\r)J_2\l(k\,\theta\sin_K\chi\r).
\nonumber
\end{eqnarray}
Here $\tau$ denotes a conformal time, 
$\chi\equiv\tau_0-\tau$ (subscripts $0$
and $LS$ denote values at the present and 
last scattering respectively),
and $P_\phi\l(k,\tau\r)$ is the power spectrum of the gravitational potential
of the matter perturbations. The window function $W\l(\chi,\chi_{LS}\r)$
is given by the expression:
\begin{equation}
W^2\l(\chi,\chi_{LS}\r)={\sin_K\l(\chi_{LS}-\chi\r)\over \sin_K \chi_{LS}},
\label{wchi}
\end{equation}
where $\sin_K \chi$ gives the distance traveled by the photon emitted
at $\tau_0-\chi$ \ie
\begin{equation}\sin_K \chi = \l
\{\begin{array}{l r} 
\displaystyle{K^{-1/2} \sin K^{1/2}\chi,}&\displaystyle{K>0;}\\ 
\displaystyle{\chi,}&\displaystyle{K=0;}\\
\displaystyle{\l(-K\r)^{1/2} \sinh \l(-K\r)^{1/2}\chi,}&\displaystyle{K<0.}
\end{array}
\r.
\label{sink}
\end{equation}

The above set of equations has been derived recently by Zaldarriaga \&\ Seljak
(1998) [see Zaldarriaga \&\ Seljak (1998) equations (7)-(10)], except
that we have set the upper limit of the $\theta$-integration in
equations (\ref{wT}-\ref{wTE}) to infinity to ensure the correct asymptotic
limit $\wtilde{C}_\ell\rightarrow C_\ell$ for $\sigma\l(\theta\r), \sigma_2\l(\theta\r) \rightarrow 0$.

The gravitational lensing correction to the CMB anisotropies depends
on the full matter power spectrum, including non-linear contributions
from small spatial scales. Previous calculations (e.g. Cole \&\
Efstathiou 1989, Seljak 1996, Zaldarriaga \&\ Seljak 1998) have
suggested that the contribution of
non-linear modes introduces only minor corrections to the
gravitational lensing contribution computed using the linear form of
the matter power spectrum. To check  whether non-linear evolution
affects our analysis we have modelled
the the non-linear corrections to the matter power spectrum using the
approach of Peacock \&\ Dodds (Peacock \&\ Dodds 1996). For values of
$\sigma_8\l(t_0\r)$ given in equation (\ref{cluster}), the Peacock-Dodds non-linear
corrections are indeed small contributions (see the bottom panels of
Figures 2a,b) and can be neglected for most purposes.

Recently Zaldarriaga and Seljak (1998) released a version of CMBFAST
that includes gravitational lensing. A comparison between our code and
theirs shows good agreement with differences smaller than $5\%$ of a
lensing generated contribution.

\begin{table}
\caption{Parameters of the cosmological models discussed in this paper.}

\begin{center}
\begin{tabular}{|c c c c c c c|}\hline
& $\omega_{\rm m}$& $\omega_{\rm b}$ & $\omega_{K}$ & $\omega_\Lambda$  & $h$  & $\sigma_8\l(t_0\r)$ \\ \hline
\multicolumn{7}{l}{$\bullet$ standard CDM model:}\\
    & $0.25$ & $0.125$ & $0.0$ & $0.0$ & $0.5$ & $0.52$ \\ \hline
\multicolumn{7}{l}{$\bullet$ class 1 of models:}\\
(a) & $0.1$ & $0.0125$ & $0.0$    & $0.15$   & $0.5$ & $0.8$  \\
(b) & $0.1$ & $0.0125$ & $0.0085$ & $0.2515$ & $0.6$ & $0.95$ \\
(c) & $0.1$ & $0.0125$ & $0.0174$ & $0.3726$ & $0.7$ & $1.1$  \\
(d) & $0.1$ & $0.0125$ & $0.0269$ & $0.5131$ & $0.8$ & $1.24$ \\ \hline
\multicolumn{7}{l}{$\bullet$ class 2 of models:}\\
(a) & $0.1$ & $0.0125$ & $0.15$   & $0.0$    & $0.5$ & $0.8$  \\
(b) & $0.1$ & $0.0125$ & $0.1756$ & $0.0844$ & $0.6$ & $0.95$ \\
(c) & $0.1$ & $0.0125$ & $0.2018$ & $0.1882$ & $0.7$ & $1.1$  \\
(d) & $0.1$ & $0.0125$ & $0.2284$ & $0.3116$ & $0.8$ & $1.24$ \\
\hline \hline 
\end{tabular}
\label{tab1a}
\end{center}
\end{table}

\begin{table*}
\caption{$\hat\chi^2$ statistics (see eqn. \ref{chi2}) 
computed for the two classes of models defined in Table 1 
and for the experimental parameters defined in
Section 3.2. The effective beam cutoffs ($\ell_{\rm beam}$) 
for the satellite missions are  given in the second row.
$\ell_{\rm max}$ denotes the number of $C_\ell$ coefficients included in the sum of equation (\ref{chi2}).
The value of $\ell_{max}$ is equal to the cut-off in $\ell$-space for
cosmic variance limited experiments and it is chosen to maximize $\hat\chi^2$
(see figure 3) for Map and Planck. The results are given for
temperature (T), polarisation (E) and cross-correlation (C) power spectra.
}

\begin{center}
\begin{tabular}{|c c c c c c c c c c c c c c c c |}\hline
& \multicolumn{9}{c}{cosmic variance limited experiment} &\multicolumn{3}{c}{MAP+} 
              &\multicolumn{3}{c}{PLANCK}\\
&             \multicolumn{3}{c}{$\ell_{max}=1000$} &
              \multicolumn{3}{c}{$\ell_{max}=2000$} & \multicolumn{3}{c}{$\ell_{max}=2750$}
              &\multicolumn{3}{c}{$\ell_{\rm beam}\sim 600$}
              &\multicolumn{3}{c}{$\ell_{\rm beam}\sim 1220$}
\\ \hline
& T & E & C & T & E & C & T & E & C & T & E & C & T & E & C
\\
\multicolumn{8}{l}{$\bullet$ standard CDM model: $\omega_{\rm m}=0.25$, $\omega_{\rm b}=0.125$}&
\multicolumn{3}{r}{$\ell_{\rm max}\simeq$ $400$ } & $400$ & $500$ & $2200$ & $1350$ & $1600$\\ 
& $0.02$ & $0.15$ & $0.08$ & $0.2$ & $2.2$ & $0.8$ & $7.3$ & $9.2$ & $3.4$
& $\simlt 10^{-2}$ & $\simlt 10^{-2}$ & $\simlt 10^{-2}$ & $0.09$ & $0.07$ & $0.1$\\ \hline
\multicolumn{8}{l}{$\bullet$ class 1 of models:  $\omega_{\rm m}=0.1$, $\omega_{\rm b}=0.0125$}
&\multicolumn{3}{r}{$\ell_{\rm max}\simeq$ $1000$ } & $600$ & $900$ & $2500$ & $1800$ & $1950$ \\
(1a) & $0.07$  & $0.9$ & $0.3$ & $0.9$ & $12.5$ & $4.2$ & $4.8$ & $33.0$ & $12.2$
& $0.02$ & $\simlt 10^{-3}$ & $\simlt 10^{-2}$ & $0.8$ & $1.5$ & $1.2$ \\
(1b) & $0.09$ & $1.2$ & $0.4$ & $1.1$ & $16.3$ & $5.5$ & $6.11$ & $42.8$ & $15.8$
& $0.02$ & $\simlt 10^{-3}$ & $\simlt 10^{-2}$ & $1.2$ & $2.2$ & $1.7$ \\
(1c) & $0.1$ & $1.6$ & $0.55$ & $1.5$ & $21.2$ & $7.1$ & $7.8$ & $55.6$ & $20.4$
& $0.03$ & $\sim 10^{-3}$ & $\sim 10^{-2}$ & $1.7$ & $3.4$ & $2.5$ \\
(1d) & $0.13$ & $2.0$ & $0.7$ & $1.8$ & $26.2$ & $8.7$ & $9.5$ & $68.3$ & $25.1$
& $0.05$ & $\sim 10^{-3}$ & $\sim 10^{-2}$& $2.3$ & $4.7$ & $3.2$ \\ \hline
\multicolumn{8}{l}{$\bullet$ class 2 of models: $\omega_{\rm m}=0.1$, $\omega_{\rm b}=0.0125$}&
\multicolumn{3}{r}{$\ell_{\rm max}\simeq$ $1000$ } & $760$ & $850$ & $2650$ &
              $2100$ & $2500$ \\
(2a) & $0.02$ & $0.5$ & $0.07$ & $0.3$ & $2.6$ & $1.3$& $0.8$ & $11.2$ & $4.7$ & 
$0.01$ & $\simlt 10^{-2}$ & $\sim 10^{-2}$ & $0.4$ & $0.95$ & $0.75$ \\
(2b) & $0.02$ & $0.7$ & $0.08$ & $0.35$ & $3.45$ & $1.7$ & $1.0$ & $14.5$ & $6.2$ & 
$0.01$ & $\simlt 10^{-2}$ & $\simlt 10^{-2}$ & $0.6$ & $1.5$ & $1.1$\\
(2c) & $0.03$ & $0.9$ & $0.1$ & $0.45$ & $4.5$ & $2.2$ & $1.3$ & $18.6$ & $7.9$
& $0.01$ & $\simlt 10^{-2}$ & $\simlt 10^{-2}$ & $0.75$ & $2.2$ & $1.6$ \\
(2d) & $0.03$ & $1.1$ & $0.13$ & $0.55$ & $5.5$ & $2.7$ & $1.6$ & $22.7$ & $9.7$ & 
$0.02$ & $\simlt 10^{-2}$ & $\simlt 10^{-2}$ & $0.9$ & $3.0$ & $2.1$   \\
\hline \hline 
\end{tabular}
\label{tab1}
\end{center}
\end{table*}

\setcounter{figure}{2}

\begin{figure}
\centerline{
\psfig{file=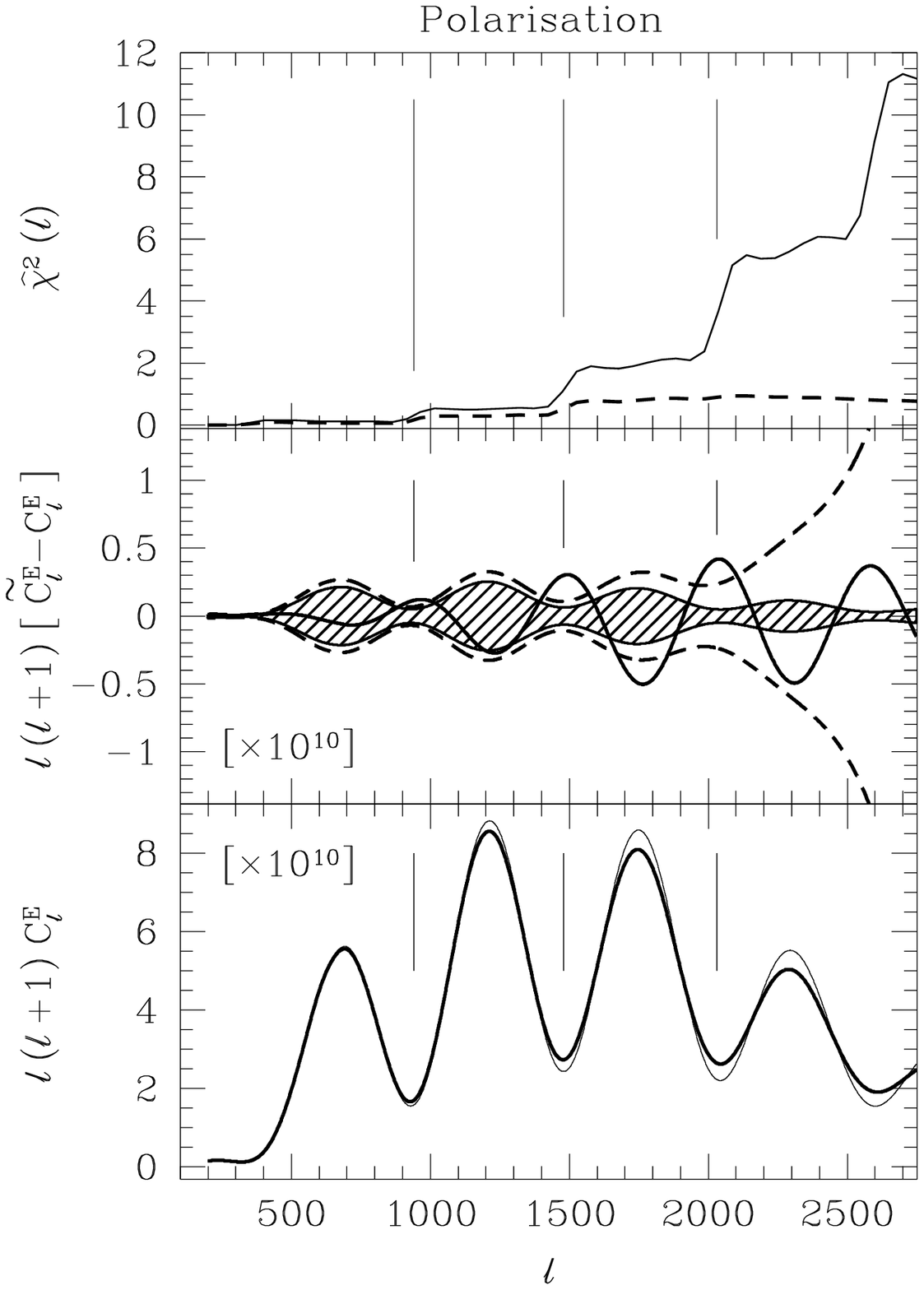,width=12cm,height=18.5cm,bbllx=3.0cm,bburx=24cm,bblly=-8.5cm,bbury=25.0cm,rheight=10cm,rwidth=5cm}
}
\caption{
The top panel shows the dependence of the reduced 
$\hat\chi^2$ (equation \ref{chi2}) on the value of
the high-$\ell$ cutoff  for a cosmic variance limited
(solid line) and a Planck-type observation (dashed line) of the
polarisation
power spectrum.  The thick oscillatory line in the middle panel
shows the gravitational lensing correction compared with the statistical
uncertainties of a cosmic variance limited experiment
(shaded region) and a Planck-like  experiment
(region delimited by dashed lines). The corresponding
linear (thick solid line) and total lensed power spectra (thin solid line)
are shown in the bottom panel. The results shown in these
figures are for model 2a with parameters listed in Table 1.
The vertical thin lines in each panel show the positions of 
minima in the linear power spectrum. The jumps in the reduced  $\hat\chi^2$ (top
panel) occur at the  minima because that is where the lensing correction
is the largest  and the cosmic variance is the smallest. }
\label{fig3}
\end{figure}

\subsection{Detectibility}

To assess whether the lensing effects are detectable, we analyse the
two classes of degenerate models illustrated in Figures \ref{fig1}
and 2.  Models of the first class (hereafter Class 1) have power
spectra which are indistinguishable from that of a spatially flat
cosmological model with $\omega_{\rm m}=0.1,$ $\omega_{\rm b}=0.0125$
and $\omega_\Lambda=0.15$. Class 2 models have power spectra that are
indistinguishable with that of an open universe with zero cosmological
constant, $\omega_{\rm m}=0.1$, $\omega_{\rm b}=0.0125$ and
$\omega_K=0.15$. The models in each class are labelled with a letter,
(a, b, c, d) in ascending order of the value of the Hubble constant
($h=0.5,0.6,0.7, 0.8$).  For all models we assume a precisely
scale-invariant spectrum of scalar adiabatic perturbations and no
contribution from tensor modes.

The parameters for the two families of models are specified in Table
1.  For comparison we have also computed results for the `standard'
CDM model [$\omega_{\rm m}=0.25$, $\omega_{\rm b}=0.0125$,
$\omega_K=\omega_\Lambda=0$, $\sigma_8(t_0)=0.52$; hereafter SCDM].  
To explore how the
results depend on the spatial resolution of the CMB observations, we
have analysed idealized examples of cosmic variance limited
observations of the power spectrum with upper multipole limits of
$\ell_{\rm max} = 1000$, $2000$ and $2750$.  Also, to assess how more
realistic observations might perform, we investigate two experimental
set-ups corresponding approximately to the MAP and Planck satellite
missions. The assumed specification are as follows: (1) the improved
MAP (\footnote{see the MAP homepage: http://map.gsfc.nasa.gov.})  best
channel parameters (hereafter MAP+) with a total power detector noise
$w_T^{-1}=2.3\times10^{-15}$ and a resolution
$\theta_{fwhm}=13.5^\prime$; (2) the $90$GHz, $150$GHz and $220$GHz of
the Planck (\footnote{see the Planck homepage:\\
http://astro.estec.esa.nl/SA-general/Projects/Planck.})
satellite: $w_{T\l(1\r)}^{-1}=2.8\times10^{-17},$
$\theta_{fwhm\l(1\r)}=16^\prime$;
$w_{T\l(2\r)}^{-1}=1.5\times10^{-17},$
$\theta_{fwhm\l(2\r)}=10^\prime$; and
$w_{T\l(3\r)}^{-1}=0.5\times10^{-17},$
$\theta_{fwhm\l(3\r)}=6.6^\prime$.  The numbers and notation follow
those of Bond \etal (1997) with $w_T^{-1}$ defined for a single channel
as the squared product of the noise level per pixel and the angular
size of a pixel.  The polarisation detector noise is assumed to be
$1.5$ and $2$ times that of the total power noise for MAP and Planck 
(see Puget \etal 1998 for a recent discussion of the polarisation
characteristics of the Planck HFI) respectively. 
In analysing both satellites, we assume that Galactic
 emission is negligible (or subtractable to high accuracy) over a fraction
$f_{sky}=0.65$ of the sky.

We assess the amplitude of the gravitational lensing corrections to
$C_\ell$ by computing the reduced $\hat\chi^2$,
\begin{equation}
\hat \chi^2  =  {1 \over \ell_{\rm max}}
\sum_{\ell=2}^{\ell_{\rm max}} { (\wtilde C_\ell - C_\ell)^2 \over (\Delta C_\ell)^2},
\label{chi2}
\end{equation}
where $\wtilde C_\ell$ is the gravitationally lensed power
spectrum computed from equation (\ref{GLcl}), $C_\ell$ is the unlensed
linear power spectrum, and $\Delta C_\ell$ is the variance of
$C_\ell$ given as,
\begin{eqnarray}
\l[\Delta C_\ell^{T,E}\r]^2&\equiv&
{ 2 \over (2\ell+1)f_{sky}}\l(C_\ell^{T,E}+w^{-1}_{T,E}b_\ell^{-2}\r)^2,\nonumber \\
\l[\Delta C_\ell^C\r]^2&\equiv& { 1 \over (2\ell+1)f_{sky}}\l[\l(C_\ell^C\r)^2\r.+\label{cosvar}\\
&+&\l.\l(C_\ell^T+w^{-1}_Tb_\ell^{-2}\r)\l(C_\ell^{E}+w^{-1}_{E}b_\ell^{-2}\r)\r],\nonumber
\end{eqnarray}
for an observation  with $N$ antennae of different
sensitivities and Gaussian beam widths $\sigma_{b\l(i\r)},$
$i=1,...,N.$ The total noise level $w_{T,E}$ in equation (10) is the
sum of the respective noise levels for each of the channels
($w_{T,E\l(i\r)}$), and an effective beam shape $b_l$ is given by
(Bond \etal 1997), $b_\ell=w^{-1}_{T,E}\sum_i w_{T,E\l(i\r)}
\exp\l[-\ell\l(\ell+1\r)\sigma^2_{b\l(i\r)}\r].$ For cosmic variance
limited experiments,  $\ell_{\rm max}$ in equation (\ref{chi2})
is equal to  the adopted cut off in an $\ell$ space. For the MAP and
Planck satellite, we choose the value $\ell_{\rm max}$  that maximizes
$\hat \chi^2$ as given in Table \ref{tab1} (see also figure 3).

A value of $\hat \chi^2$ of order unity signifies that the lensing
distortion is detectable in principle by an experiment.
In this Section we use this criterion as a rough measure
of the detectability of gravitational lensing in the CMB.  However, if
we parameterize $C_\ell$ by $N$ cosmological parameters, then values
of $\hat \chi^2$ as low as $\sim N/\ell_{\rm max}$ can lead to significant
differences between estimated parameters. The effect of lensing on
cosmological parameters is discussed in more detail in Section 4.3.

Values of $\hat \chi^2$ are listed in Table \ref{tab1} for various
experimental setups.  
For cosmic variance limited observations, the
reduced $\hat\chi^2$ values are approximately proportional to the fourth
power of the mass spectrum normalisation amplitude $\sigma_8\l(t_0\r)$.

\begin{figure*}
\centerline{
\psfig{file=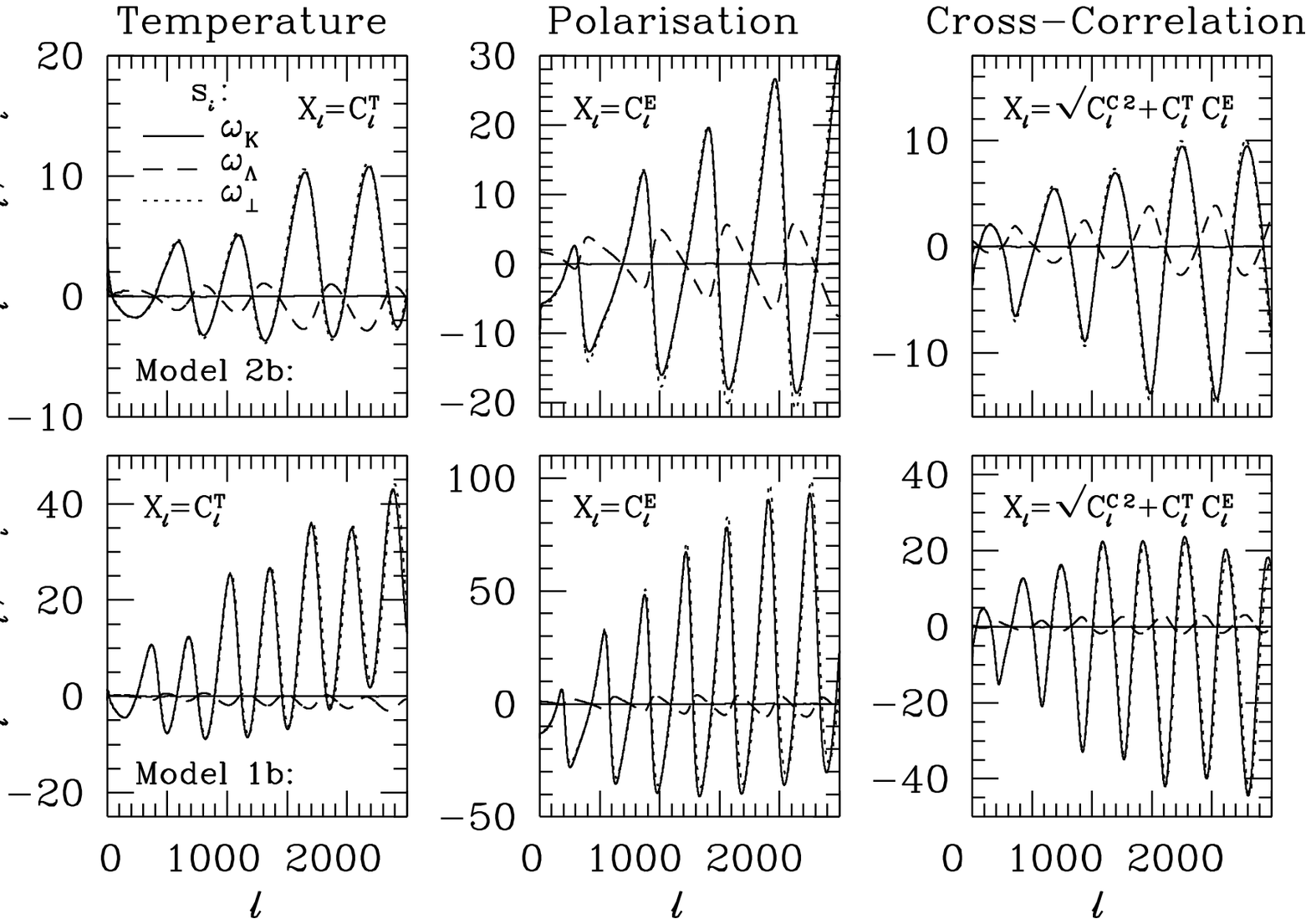,width=17cm,rheight=11cm,rwidth=14cm}
}
\caption{
The derivatives of the linear power spectrum of
temperature (left panels), polarisation (middle panels) and cross-correlation
(right panels)
with respect to $\omega_K$ (solid lines), $\omega_\Lambda$ (dashed),
and $\omega_\perp$ (dotted). Model 2b is shown in the upper
figures and model 1b in the lower figures. The nearly horizontal lines
in each panel show the numerically
computed derivative along the degeneracy direction [\ie with respect
to $\omega_\parallel$ see eqn.(\ref{newpars})].
}
\label{fig4}
\end{figure*}

\begin{figure*}
\centerline{
\psfig{file=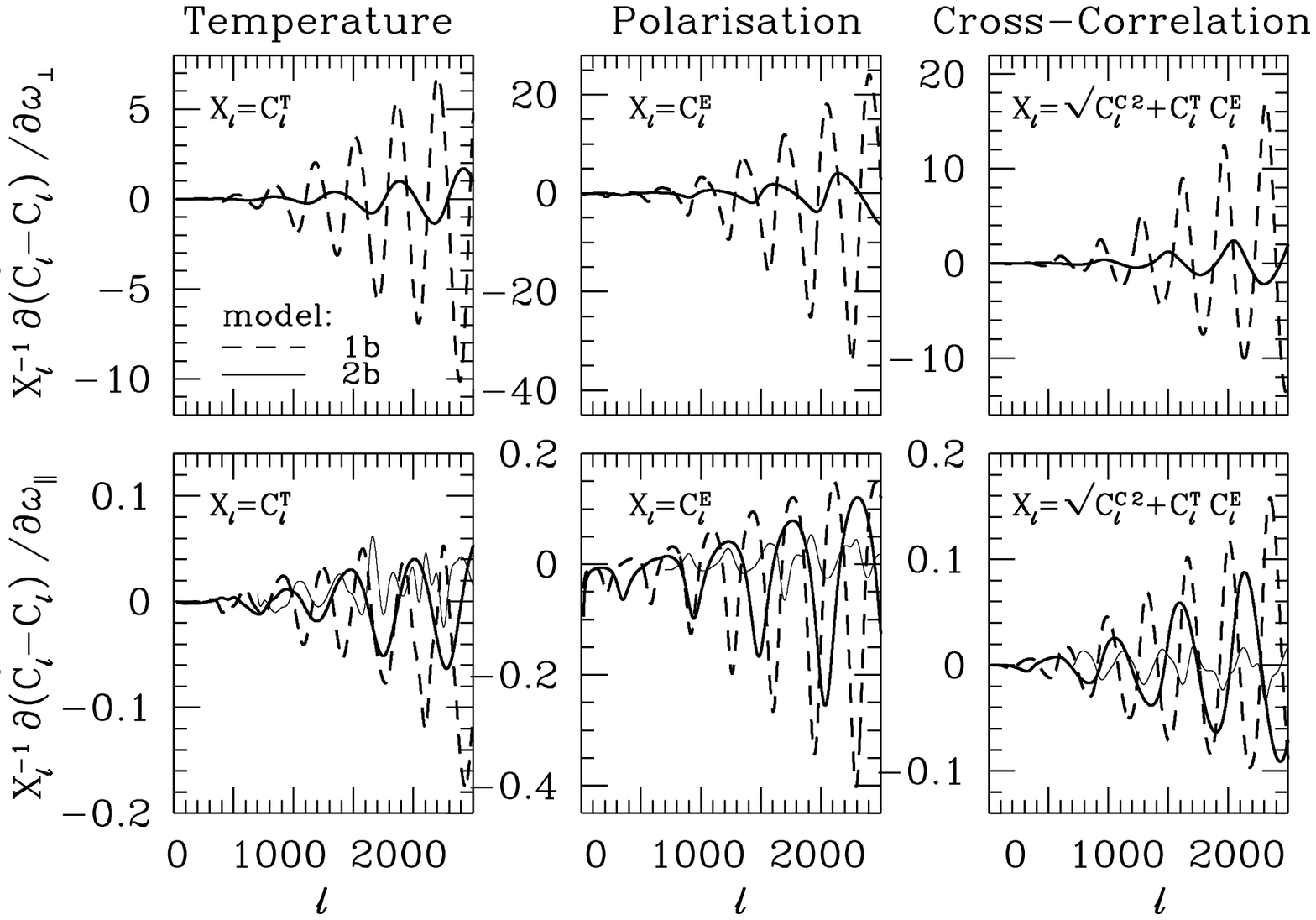,width=17cm,rheight=11cm,rwidth=14cm}
}
\caption{
The derivatives of the gravitational lensing
contribution to temperature (left panels), polarisation (middle
panels) and cross-correlation (right panels)
with respect to $\omega_\perp$ 
(upper row) and $\omega_\parallel$ (lower row) for 
models 1b and 2b. 
Dashed  oscillatory lines show the derivatives for the spatially flat model
(1b) and solid thick lines show the derivatives for the open model (2b).
The thin solid lines in the lower panels show the linear power spectrum
derivatives with respect to $\omega_\parallel,$ as plotted in Figure 
\ref{fig4}.}
\label{fig5}
\end{figure*}

For the normalisation required by the observed present day cluster
abundance (equation \ref{cluster}), the gravitational lensing contributions to the
CMB power spectra are small but not negligible at the sensitivities of
a Planck-type experiment. Since the lensing corrections depend
strongly on the amplitude of the mass fluctuations at the relatively
high redshifts, the
detectability of lensing is sensitive to the parameters of the target
model. Thus, the reduced $\hat\chi^2$ values for the standard CDM model
listed in Table \ref{tab1} for a Planck-type experiment are much lower than
those of the $\Lambda$-dominated and open models listed in the table,
which have higher values of $\sigma_8$ and lower perturbation growth
rates at the present. Gravitational lensing modifies
the `damping tail' of the temperature power spectrum at high
multipoles (Metcalf \&\ Silk 1998) and this effect makes a significant
contribution to $\hat\chi^2$ in cosmic variance limited experiments
probing $\ell_{\rm max} \simgt 2000$. The detectability of lensing is
thus sensitive to the multipole range probed by an experiment and to
the cosmological angle-distance relation; the effect of lensing is
less easy to detect in a negatively curved universe since the damping
tail is pushed to higher multipoles and power spectrum peaks are broader
than in a spatially flat universe.
This is why the $\hat \chi^2$ values in Table \ref{tab1}
for the $\omega_K = 0.15$ model are lower than those of the $\omega_K=0$
model at fixed $\ell_{\rm max}$.

The effects of gravitational lensing on the polarisation power
spectrum are even more significant than for the temperature power
spectrum, producing significant distortions at $\ell \sim 1000$. At first
sight the high values of $\hat \chi^2$ for polarisation might seem
surprising. They arise because the polarisation power spectrum has
sharper minima and peaks compared to the temperature power spectrum
(\cf Figures 2a,b) and is therefore more sensitive to the
gravitational lensing convolution (equation 5).  This build up of
$\hat \chi^2$ with multipole for polarisation is illustrated in Figure~3.

However, as the amplitude of the polarisation power spectrum is almost
two orders of magnitude lower than that of the temperature
fluctuations, an experiment with high sensitivity to polarisation, in
addition to high angular resolution, is required to detect the lensing
contribution.  Indeed, for a MAP type mission, the results in Table
\ref{tab1} show that gravitational lensing is difficult to detect in
either temperature or polarisation spectra. However, for Planck the
lensing effects might be significant in both power spectra and, despite the
lower amplitude, are usually more easily detected in  polarisation than in
the temperature signal. 
In a Planck-like (or cosmic variance limited) experiment,
the lensing contribution is measurable also in the cross-correlation power
spectrum, though  less easily than in the polarisation spectrum.

\subsection{The dependence of gravitational lensing 
contributions on cosmological parameters}

The dependence of the gravitational lensing contribution on
cosmological parameters can be quantified by computing the derivatives
of the lensed power spectra with respect to the cosmological parameter
of interest. However, this presents a difficult numerical problem,
because the lensing contribution is small and hence the derivatives can
be easily swamped by numerical errors.  In particular, the direct
finite differencing scheme used to compute linear power spectrum
derivatives (Bond \etal 1997, Zaldarriaga \etal 1997;
Eisenstein \etal 1998,  see also Section 4.2), cannot be applied to
the lensed case given
the typical numerical errors of $\sim 1 \%$ in CMB Boltzmann codes.
Instead we have applied two different semi-analytical approaches based
on numerical derivatives of the matter power spectrum rather than
those of $\wtilde C_\ell$.

The derivative with respect to a cosmological parameter 
$s_i$ can be expressed as ($I=T,E,C$),
\begin{equation}
{\partial \wtilde C_\ell^I \over \partial s_i}={\partial C_\ell^I \over
\partial s_i}+\sum_{\ell'}\l[{\partial W^I_{\ell\ell'} \over \partial s_i} C^I_{\ell}+
W^I_{\ell\ell'}{\partial C_\ell^I \over \partial s_i}\r]
\label{clder}
\end{equation}
where computing derivatives of the window functions ${\cal W}^I$ 
involves computations of the derivatives of the photon path
dispersions $\sigma\l(\theta\r)$ and $\sigma_2\l(\theta\r)$.

In our first approach, derivatives of
$\sigma^2\l(\theta\r)$, $\sigma_2^2\l(\theta\r)$ 
and $C_\ell$ are computed numerically by finite
differencing. In the second method,  only the 
derivative of $C_\ell$ is computed by finite differencing, and the
derivatives of both $\sigma^2\l(\theta\r)$ and $\sigma_2^2\l(\theta\r)$ 
are calculated using numerically precomputed
derivatives of the gravitational potential power 
spectrum $P_{\phi}\l(k,\eta\r)$. For example,
 derivatives of $\sigma^2\l(\theta\r)$
are obtained through the formula,
\begin{eqnarray}
{\partial \sigma^2\l(\theta\r)\over \partial s_i}&=&16\pi^2\int_0^\infty dk
k^3 \int_{a_{LS}}^{a_{0}} da \times \label{dsig}\\
&\times&\l\{{\partial P_{\phi}\l(k,a\r)\over \partial
s_i}\r.
 W^2\l[1-J_0\l(k\theta\sin_K\chi\r)\r]
{\partial \chi\over \partial a}\nonumber \\
&+&\l.P_{\phi}\l(k,a\r)\r. \l.{\partial\over \partial s_i}
\l[W^2\l[1-J_0\l(k\theta\sin_K\chi\r)\r]{\partial 
\chi\over \partial a}\r]\r\}, \nonumber
\end{eqnarray}
derived from equation (\ref{sigmath}). 
An analogous expression can be written  for 
$\sigma_2^2\l(\theta\r)$.
To compute the derivative of $P_{\phi}$ at any given value of the scale
factor $a$, we assumed that $P_{\phi}$ grows according to linear
theory, $P_{\phi}(k,a)=D^2(a,a_{LS})P_{\phi}(k,a_{LS})$, where
$D(a, a_{LS})$ is the linear growth factor and depends only
the scale factor and cosmological parameters.
The growth factor has been computed directly by numerical
integration of the set of linearized equations of motion (assuming
Newtonian dynamics and a pressureless fluid).
The derivatives of the growth factor have also been computed 
by integrating a set of derivatives of the equations of
motion with respect to cosmological parameters.

For most target models that we have investigated, we find good
agreement between the derivatives computed from the two methods
described above. The second method is the more accurate, especially
for cases where the parameter dependences of the lensed spectrum are
weak, e.g. the derivative of $\wtilde C_\ell$ with respect to
the residual optical depth for Thomson scattering.  However, for most
of the cosmological parameters, and in particular the derivatives with
respect to curvature $\omega_K$ and $\omega_\Lambda$, both methods
fared equally well and we have chosen to use the first one since it is
more computationally efficient.

\begin{figure*}
\centerline{
\psfig{file=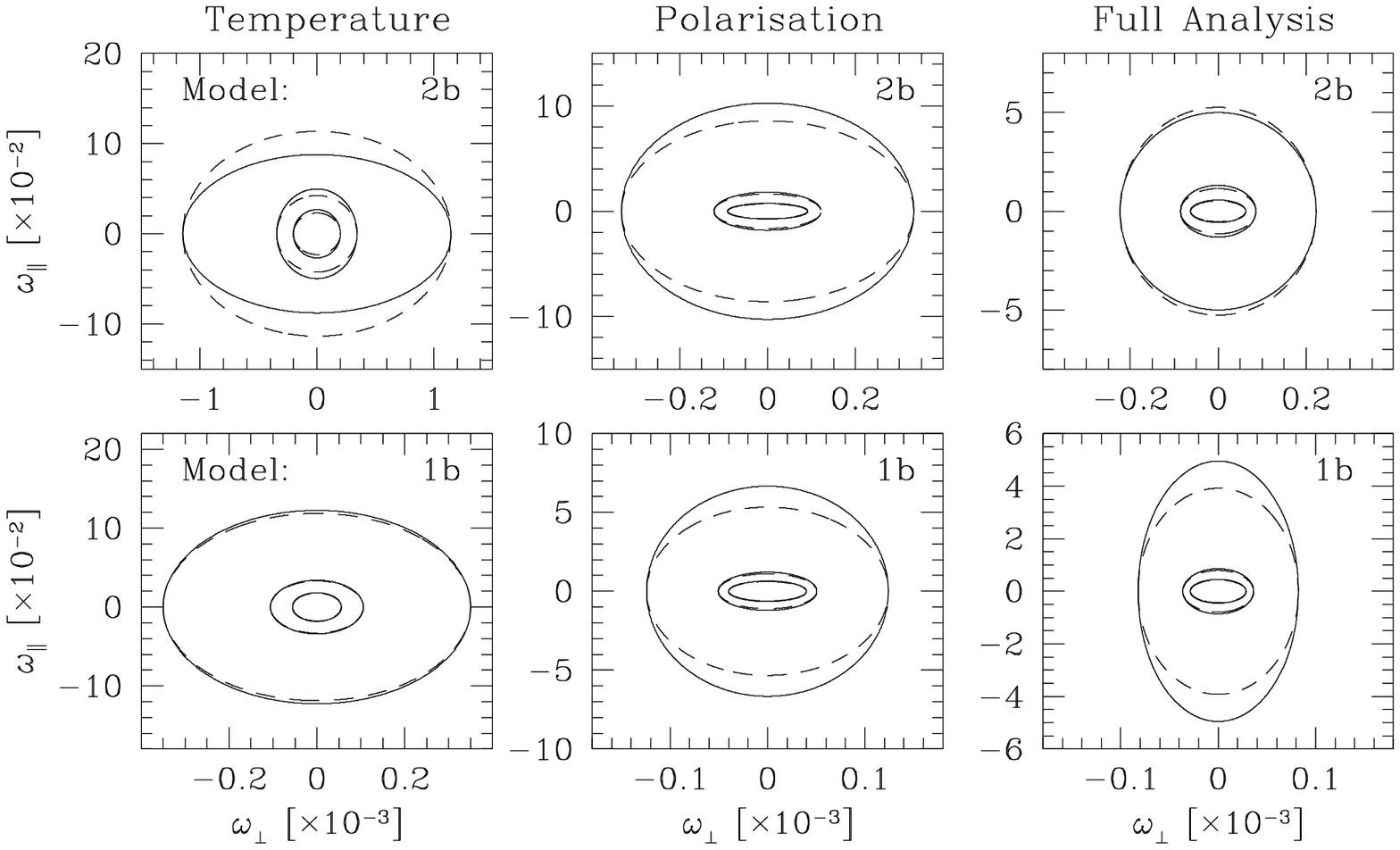,width=16.5cm,rheight=10.0cm} 
}
\caption{
The $2\sigma$ constraints on the parameters $\omega_\perp$ and
$\omega_\parallel$ (see eqn. (\ref{newpars}) for a definition)
derived from the Fisher matrix analysis described in the text.
The results shown are for  models  1b (bottom row) and 2b (upper
row) and for three idealized cosmic variance limited experiments with 
$\ell_{\rm max}=1000$ (giving the weakest constraints)
$2000$ and  $2750$ (giving the strongest constraints).
The solid contours are derived from our numerical computations
of the $C_\ell$ derivatives. The dashed lines
assume vanishing linear $C_\ell$  derivatives along 
the degeneracy directions for 
$\ell  \simgt 200$ (\ie imposing the
condition that the  geometrical degeneracy is exact at
high multipoles). The differences between the solid and
dashed contours provide an estimate of the numerical errors
in the computations of the  Fisher matrix.
}

\label{fig6}
\end{figure*}

The derivatives of the lensing corrections with respect to the
parameters $\omega_K$, $\omega_\Lambda$ and $\omega_\parallel$ are
shown in Figure \ref{fig5} for models 1b and 2b. These figures can be
compared directly to Figure \ref{fig4} showing derivatives of the
linear (unlensed) power spectra for these models.

Figure \ref{fig5} shows  that the gravitational lensing derivatives are 
small. This is because the matter power spectrum, on which the
lensing correction depends via equations (\ref{GLcl}-\ref{sigmath2}), is 
insensitive to many of the cosmological parameters. Furthermore,
the main contribution to lensing comes from  matter at relatively high 
redshifts, when the growth rates of perturbations are still largely
insensitive to the cosmological model. Nevertheless, as we will show
in the next Section, these weak dependences on cosmological parameters
are potentially detectable and could break the geometrical degeneracy
of linear CMB power spectra.

\section{CMB constraints on curvature and the cosmological constant}

\subsection{The Fisher matrix formalism}

For a vector $\bar{x}$ of N Gaussian variables depending upon $s_i$
($i=1,...,n,$) parameters, an $n\times n$ Fisher matrix is defined as
follows (e.g. Kendall \&\ Stuart 1967, Tegmark, Taylor \&\ Heavens 1997) 
\begin{equation}
F_{ij}= {\partial \bar{x} \over \partial s_i}^T {\hat M}^{-1} {\partial \bar{x} \over \partial s_j},
\label{fisher1}
\end{equation}
where $\hat M$ denotes the correlation matrix of $\bar{x}.$ For
independent
measurements of the
temperature or polarisation power spectra, $\bar x$ is a vector
of the respective power spectrum coefficients and 
$\hat M$ is diagonal with dispersions given by equation (14).
The Fisher matrix in this case reads,
\begin{equation}
F_{ij}^{T,E}=\sum_{\ell=2}^{\ell_{\rm max}} {\partial C_\ell^{T,E}\over \partial s_i}\,
{\Delta {C_\ell}^{T,E}}^{-2}\, {\partial C_\ell^{T,E}\over \partial s_j}.
\label{fisher2}
\end{equation}
For a simultaneous analysis of all three power spectra
we choose 
${\rm Transpose}\l({\bar x}\r) = [C_2^T,C_2^E,C_2^C,...,$
$C_\ell^T,C_\ell^E,C_\ell^C...],$ ensuring a block
diagonal form of the $\hat M$ matrix with $3\times3$ diagonal
submatrices ($\equiv \hat M_\ell$) describing correlations of the $\l[C_\ell^T,C_\ell^E,C_\ell^C\r]$ vector. 
The appropriate Fisher matrix can be written  as (Zaldarriaga \etal
1997, Kamionkowski, Kosowsky \&\ Stebbins 1997),
\begin{equation}
F_{ij}^{T+E}=\sum_{\ell=2}^{\ell_{\rm max}} \sum_{I,J=\l\{T,E,C\r\}} {\partial C_\ell^{I}\over \partial s_i}
\l[{M_\ell}^{-1}\r]_{IJ} {\partial C_\ell^J\over \partial s_j},
\label{fisher3}
\end{equation}
where $\hat{M}_{\ell}$ is given by,
\begin{eqnarray}
\hat{M}_\ell^{TT}&=&{2f_{sky}^{-1}\over2\ell+1}\,{\l(C_\ell^T+w_T^{-1}b_\ell^{-2}\r)}^2 \nonumber\\
\hat{M}_\ell^{TE}&=&{2f_{sky}^{-1}\over2\ell+1}\,{C_\ell^C}^2\nonumber\\
\hat{M}_\ell^{TC}&=&{2f_{sky}^{-1}\over2\ell+1}\,C^C_\ell\l(C^T_\ell+w_T^{-1}b_\ell^{-2}\r)\label{Melem}\\
\hat{M}_\ell^{EE}&=&{2f_{sky}^{-1}\over2\ell+1}\,{\l(C_\ell^E+w_E^{-1}b_\ell^{-2}\r)}^2\nonumber\\
\hat{M}_\ell^{EC}&=&{2f_{sky}^{-1}\over2\ell+1}\,C^C_\ell\l(C^E_\ell+w_E^{-1}b_\ell^{-2}\r)\nonumber\\
\hat{M}_\ell^{CC}&=&{f_{sky}^{-1}\over2\ell+1}\,\l[{C^C_\ell}^2+\l(C^T_\ell+w_T^{-1}b_\ell^{-2}\r)\l(C^E_\ell+w_E^{-1}b_\ell^{-2}\r)\r]\nonumber
\label{twodcorell}
\end{eqnarray}
Finally, the inverse of the Fisher matrix $F_{ij}$ is interpreted as
the covariance matrix of the small parameter deviations ($\equiv
\delta s_i$) from their target values. In particular, the $i^{th}$
diagonal element of its inverse provides a lower bound on the standard
error of the corresponding $s_i$ parameter.

\begin{table}
\caption{$68.5\%$ (`$1\sigma$') uncertainties (multiplied by a factor
of $100$) for
$\omega_K$ and $\omega_\Lambda$ determinations (with all other parameters fixed) 
obtained through the Fisher matrix analysis of the 
CMB power spectrum with gravitational
effects included.  The results of the analysis based on either
temperature or polarisation information are denoted by $T$ and $E$, 
respectively,
while the results of the full analysis are denoted as $TEC$.
The table lists results  for the  $h=0.5$ models only.
} 

\begin{tabular}{|l c c c c c c|}\hline 
Experiment \ \ \ \ \ & \multicolumn{3}{c}{$\delta\omega_K$ $\l[\times 10^{-2}\r]$,} & 
\multicolumn{3}{c}{$\delta\omega_\Lambda$ $\l[\times 10^{-2}\r]$} \\
 & $T$ & $E$ & $TEC$ & $T$ & $E$ & $TEC$ \\
\hline \hline 
\multicolumn{7}{l}{$\bullet$ standard CDM model}\\
\multicolumn{7}{c}{($\omega_{\rm m}=0.25$, $\omega_{\rm b}=0.0125,$
$\omega_K=0$, $\omega_\Lambda=0$, $\sigma_8\l(t_0\r)=0.52$)}\\
$l_{max}=1000$ & $2.1$  & $0.8$  & $0.4$  & $14.0$ & $5.5$  & $2.8$  \\
$l_{max}=2000$ & $0.5$  & $0.15$ & $0.1$  & $3.25$ & $1.1$  & $0.7$ \\
$l_{max}=2750$ & $0.15$ & $0.075$& $0.045$& $0.85$ & $0.5$  & $0.3$ \\
Planck         & $1.1$  & $2.6$  & $0.8$  & $7.4$  & $17.1$ & $5.5$ \\
\hline
\multicolumn{7}{l}{$\bullet$ model 1a:}\\
\multicolumn{7}{c}{($\omega_{\rm m}=0.1$, $\omega_{\rm b}=0.0125,$
$\omega_K=0$, $\omega_\Lambda=0.15$, $\sigma_8\l(t_0\r)=0.8$)}\\
$l_{max}=1000$ & $0.45$ & $0.2$   & $0.15$  & $5.3$  & $2.3$  & $1.5$  \\
$l_{max}=2000$ & $0.15$ & $0.045$ & $0.03$  & $1.5$  & $0.5$  & $0.35$ \\
$l_{max}=2750$ & $0.075$& $0.025$ & $0.02$  & $0.8$  & $0.3$  & $0.2$ \\
Planck         & $0.2$  & $0.2$   & $0.1$   & $2.1$  & $2.4$  & $1.2$ \\
\hline
\multicolumn{7}{l}{$\bullet$ model 2a:}\\
\multicolumn{7}{c}{($\omega_{\rm m}=0.1$, $\omega_{\rm b}=0.0125,$
$\omega_K=0.15$, $\omega_\Lambda=0$, $\sigma_8\l(t_0\r)=0.8$)}\\
$l_{max}=1000$ & $1.2$  & $1.1$  & $0.6$  & $3.5$  & $3.5$  & $1.7$ \\
$l_{max}=2000$ & $0.55$ & $0.2$  & $0.15$ & $1.6$  & $0.65$ & $0.4$ \\
$l_{max}=2750$ & $0.3$  & $0.09$ & $0.065$& $0.9$  & $0.27$ & $0.2$ \\
Planck         & $0.5$  & $0.65$ & $0.3$  & $1.5$  & $2.0$  & $0.9$  \\
\hline 

\hline 
\end{tabular}
\label{results}

\end{table}

\subsection{Numerical computations}

The linear power spectra were computed using the CMBFAST code of
Seljak \&\ Zaldarriaga (1996), and their derivatives with respect to 
cosmological parameters were obtained by a direct finite differencing scheme.
To improve the numerical accuracy, we
used much finer grids in wavenumber and multipole and shorter
integration timesteps than the default CMBFAST values.  The importance
of accurate derivatives of the CMB power spectrum in the computation
of the Fisher matrix has been discussed recently by Bond \etal (1997),
Efstathiou \&\ Bond (1998) and Eisenstein \etal (1998). In
particular, numerical errors in the derivatives required for the
Fisher matrix can easily lead to a 
spurious breaking of the geometrical degeneracy.  An accurate
determination of the Fisher matrix is especially important in the
analysis described here, because we want to investigate whether a real
physical effect, rather than numerical errors, can break the
geometrical degeneracy.

The accuracy of CMBFAST was tested by measuring derivatives of linear
power spectra along the geometrical degeneracy line (\ie with respect to
$\omega_\parallel$) as shown by the thin solid lines in the lower
panels of Figure \ref{fig5}. The residuals are significantly 
smaller than the
derivatives of the lensing corrections. As we will demonstrate in
the next section, these errors do not affect the 
Fisher matrix coefficients significantly when
lensing is included. Also the results of Fisher matrix analysis do not
depend on the choice of the initial parametrization demonstrating that
our linear power spectrum derivatives are not significantly affected by
numerical errors.

\begin{figure}
\resizebox{\columnwidth}{!}{\includegraphics{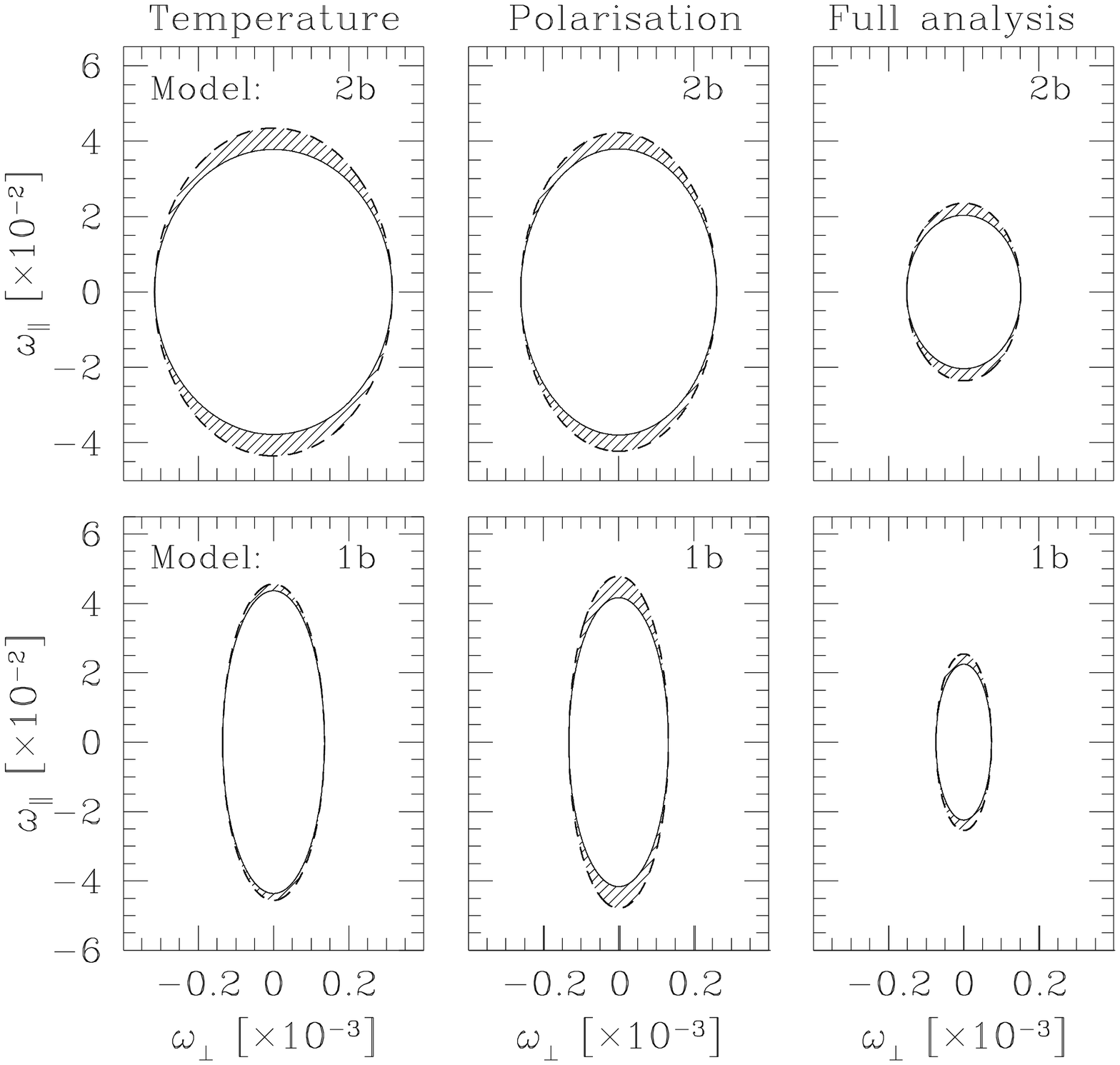}}

\caption{
$2\sigma$ contours in $\l(\omega_\perp,\omega_\parallel\r)$ plane for a Planck-type experiment. The width 
of the shaded regions shows the effect of the numerical inaccuracies on our final results.
}
\label{fig7}
\end{figure}
\subsection{Fisher matrix analysis of CMB spectra}

To analyse the degree to which the degeneracy can be broken by the
lensing contribution we compute the $C_\ell$ derivatives and hence the
Fisher matrix of the power spectra including the gravitational lensing
contribution. We assume that the lensed CMB anisotropies are
Gaussian. On the angular scales probed by MAP and Planck, the
non-Gaussianities introduced by gravitational lensing should be
negligible (see Bernardeau 1997). We first analyse the effects of
gravitational lensing in the two-dimensional space defined by
$\omega_K$ and $\omega_\Lambda$, to show that lensing does break the
geometrical degeneracy if all other parameters are known. We then
analyse a more realistic example varying six
cosmological parameters.

 \begin{figure*}
\centerline{
\psfig{file=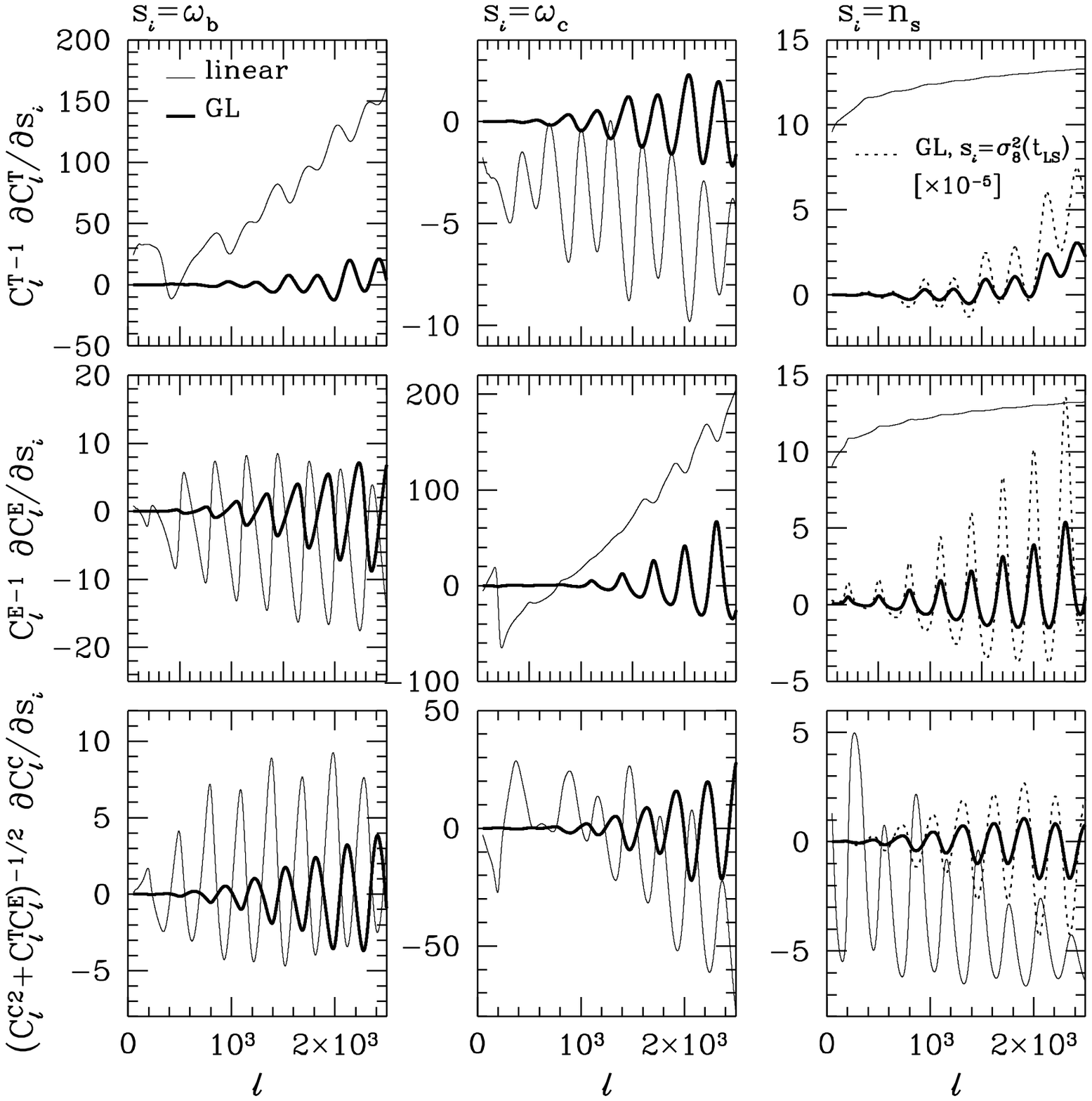,width=17.5cm,rheight=16cm}
}
\caption{
Examples of linear power spectrum and gravitational lensing contribution derivatives 
for the SCDM model (see Table 1) with respect
to three cosmological parameters. The solid thin line in each panel
shows the linear power spectrum derivative, while thick lines  show the
derivative of the lensing contribution (with respect to the 
parameter given at the top of the picture). In addition the
gravitational lensing contribution derivative with respect to an
amplitude [defined as $\sigma_8^2\l(t_{LS}\r)$ and expressed in units
$10^{-5}$] is shown as dotted lines in the rightmost panels.
}
\label{fig8}
\end{figure*}

\subsubsection{Variations of $\omega_K$ and $\omega_\Lambda$}

The Fisher matrix results restricting to the two parameters $\omega_K$
and $\omega_\Lambda$ are presented in Table 3  and in Figures \ref{fig6}
\&\ \ref{fig7}. In these Figures we plot $2\sigma$ error ellipses in
the plane $\l(\omega_\parallel, \omega_\perp\r)$ defined as in
equation (\ref{newpars}), \ie in the directions parallel and
perpendicular to the geometrical degeneracy direction. In each case,
the solid contours were derived by computing the numerical derivatives
of the linear and lensing corrections with respect to both parameters
(as shown in Figures \ref{fig4} and \ref{fig5}). The dashed contours
were computed by setting the derivative of the linear power spectra
along the degeneracy direction to zero at multipoles $\ell \simgt
200$. The differences between the solid and dashed lines thus provide
an indication of the effects of numerical errors on the Fisher matrix
analysis. For both temperature and polarisation anisotropies,
gravitational lensing {\it breaks the geometrical degeneracy} and
leads to useful constraints on $\omega_\parallel$.  For example, for a
Planck-type experiment and the two models shown in Figure 7 [models 1b
and 2b, normalised to have $\sigma_8(t_0) = 0.95$], the inclusion of
gravitational lensing leads to a $2\sigma$ error of $\omega_\parallel
\approx 0.05$ and an even a stricter constraint on $\omega_\Lambda.$
The constraints are less strong for models with lower amplitude at the
high redshifts (as a result of the normalisation of the present day
matter fluctuations or/and their higher growth rates) and hence
smaller CMB lensing contributions. In the case of the standard CDM
model listed in Table 3, the $1\sigma$ constraints on $\omega_\Lambda$
and $\omega_K$ when polarisation and temperature information are
combined are $\delta \omega_\Lambda \approx 0.05$ and $\delta \omega_K \approx
0.008$. These constraints are considerably worse than those for models
$1a$ and $2a$ listed in Table 3, which have a higher amplitude of
mass fluctuations. Nevertheless, even
for the standard CDM model, the constraints on $\omega_\Lambda$ and
$\omega_K$ are still  interesting.

\begin{table*}
\caption{Forecasted accuracy (`$1\sigma$') of the determination of the selected cosmological parameters
as obtained through Fisher matrix analysis of the six dimensional parameters space with additional external constraints 
imposed within a 2DIM $\l(\omega_K,\omega_\Lambda\r)$ plane and a gravitational lensing effect either excluded 
(\ie linear radiation power spectrum only) or included as indicated. 
Presented results are for a Planck-like experiment measuring the
$C^T$, $C^E$ and $C^C$ power spectra (see text for details)}

\begin{center}
\begin{tabular}{|l c c c c c c c|} 
\hline\hline
& GL     & $\delta \omega_b$ & $\delta \omega_c$ & $\delta \omega_K$ & $\delta \omega_\Lambda$  & $\delta n_s$ & $ {\displaystyle{\delta \sigma_8\l(t_{LS}\r)}\over \displaystyle{\sigma_8\l(t_{LS}\r)}}$ \smallskip\\
& contr. & $\l[\times 10^{-5}\r]$ & $\l[\times 10^{-3}\r]$ & $\l[\times 10^{-2}\r]$ & $\l[\times 10^{-2}\r]$ & $\l[\times 10^{-3}\r]$ & $\l[\times 10^{-2}\r]$\\
\hline\hline
$\bullet$ SCDM: & & $\omega_b=0.0125$ & $\omega_c=0.2375$ & $\omega_K=0$ & $ \omega_\Lambda=0$ & $n_s=1$ & $\sigma_8\l(t_0\r)=0.52$\\
\ \ full 6D case:             & yes & $6.2$ & $2.6$ & $1.0$   & $6.3$   & $2.9$ & $1.9$ \\
\ \ 2D case:                  & yes & $-$   & $-$   & $0.8$   & $5.5$   & $-$   & $-$  \\
\ \ $\omega_\parallel$ fixed: & yes & $6.1$ & $2.3$ & $0.1$   & $0.014$ & $2.9$ & $1.8$ \\
                              & no  & $5.9$ & $2.4$ & $0.1$   & $0.015$ & $3.1$ & $2.0$ \\
\ \ $\omega_\Lambda$ fixed:   & yes & $5.6$ & $1.3$ & $0.06$  & $-$     & $2.8$ & $1.7$ \\
                              & no  & $5.3$ & $1.3$ & $0.06$  & $-$     & $3.0$ & $1.8$ \\
\ \ $\omega_K$ fixed:         & yes & $6.1$ & $2.5$ & $-$     & $0.7$   & $2.9$ & $1.8$ \\
                              & no  & $6.0$ & $2.7$ & $-$     & $0.7$   & $3.1$ & $2.0$ \\
\hline
$\bullet$ model 1a: & & $\omega_b=0.0125$ & $\omega_c=0.0875$ & $\omega_K=0$ & $ \omega_\Lambda=0.15$ & $n_s=1$ & $\sigma_8\l(t_0\r)=0.8$\\ 
\ \ full 6D case:             & yes & $4.2$ & $0.4$   & $0.3$   & $3.4$    & $2.1$ & $1.3$ \\
\ \ 2D case:                  & yes & $-$   & $-$     & $0.1$   & $1.2$    & $-$   & $-$  \\
\ \ $\omega_\parallel$ fixed: & yes & $4.2$ & $0.24$  & $0.01$  & $0.001$  & $1.8$ & $1.0$ \\
                              & no  & $4.1$ & $0.17$  & $0.009$ & $0.0008$ & $2.3$ & $1.3$ \\
\ \ $\omega_\Lambda$ fixed:   & yes & $4.2$ & $0.2$   & $0.01$  & $-$      & $1.9$ & $1.0$  \\
                              & no  & $4.1$ & $0.15$  & $0.008$ & $-$      & $2.3$ & $1.3$ \\
\ \ $\omega_K$ fixed:         & yes & $4.5$ & $0.5$   & $-$     & $0.3$    & $1.5$ & $0.8$  \\
                              & no  & $4.2$ & $0.7$   & $-$     & $0.4$    & $2.6$ & $1.8$  \\
\hline
$\bullet$ model 2a: & & $\omega_b=0.0125$ & $\omega_c=0.0875$ & $\omega_K=0.15$ & $ \omega_\Lambda=0$ & $n_s=1$ & $\sigma_8\l(t_0\r)=0.8$\\
\ \ full 6D case:             & yes & $4.0$ & $0.4$  & $0.7$   & $1.9$   & $1.5$ & $0.7$  \\
\ \ 2D case:                  & yes & $-$   & $-$    & $0.3$   & $0.9$   & $-$   & $-$   \\
\ \ $\omega_\parallel$ fixed: & yes & $3.9$ & $0.24$ & $0.06$  & $0.02$  & $1.4$ & $0.6$   \\
                              & no  & $3.6$ & $0.23$ & $0.06$  & $0.02$  & $1.6$ & $0.8$  \\
\ \ $\omega_\Lambda$ fixed:   & yes & $3.8$ & $0.2$  & $0.05$  & $-$     & $1.4$ & $0.6$  \\
                              & no  & $3.6$ & $0.2$  & $0.05$  & $-$     & $1.6$ & $0.7$  \\
\ \ $\omega_K$ fixed:         & yes & $3.9$ & $0.2$  & $-$     & $0.2$   & $1.4$ & $0.6$  \\
                              & no  & $3.6$ & $0.2$  & $-$     & $0.2$   & $1.6$ & $0.8$  \\
\hline
\end{tabular}

\label{multidim}
\end{center}
\end{table*}

\subsubsection{Variations of  6 cosmological parameters}

The results of the previous section show that a Planck-type experiment
is sensitive to gravitational lensing of the CMB and that
gravitational lensing can break the geometrical degeneracy. There is,
however, a possibility that the lensing contributions to the CMB
anisotropies can be mimicked by variations in other cosmological
parameters thereby defining a new degeneracy direction. It is unlikely
that such a degeneracy would be perfect, but it is possible that the
the total power spectra including lensing corrections are partially
degenerate with respect to variations of other parameters, 
limiting their usefulness as a diagnostic of the geometry of the
Universe. To examine this possibility we have therefore computed the
Fisher matrix for a more realistic six parameter space defined by the following
cosmological parameters: $\omega_K$, $\omega_\Lambda$, $\omega_{\rm
b}$, $\omega_{\rm c}$, the scalar spectral index $n_s$ and the
amplitude $\sigma_8\l(t_{LS}\r)$ of the mass fluctuations at the time
of recombination. We have explicitly ignored a tensor component since any
tensor component would have a negligible amplitude at the high multipoles
at which gravitational lensing of the CMB is significant. 

The results of the Fisher matrix analysis for a Planck-type experiment
are listed in Table \ref{multidim}. The errors on $\omega_\Lambda$ and $\omega_K$ 
for the standard CDM model in the six parameter example are usually
larger than those of the idealized two parameter example described in the previous
subsection. For example for models $1a$ ($\Lambda$-dominated, spatially flat universe)
and $2a$ ($\Lambda=0$, open universe), the errors on $\omega_\Lambda$ and
$\omega_K$ in the six-dimensional case are about $2-3$ times those derived
for the two-dimensional case. However, no new near-degeneracy is
found suggesting that it is feasible to separate the geometrical dependence
of the lensing contributions of the CMB power spectra from the changes 
caused by varying other cosmological parameters.

\section{Summary}

Observations of the CMB anisotropies promise a dramatic improvement in our
knowledge of the formation of cosmic structure and  of the values of
fundamental cosmological parameters that define our Universe.  According
to linear perturbation theory, however, there exists a near exact
geometrical degeneracy that makes it nearly impossible 
to disentangle the values
of $\Omega_K$ and $\Omega_\Lambda$ from observations of the CMB
anisotropies alone. In reality, the CMB temperature and polarisation
anisotropies will be modified by gravitational lensing caused by the
irregular distribution of matter between us and the last scattering
surface. The effects of gravitational lensing, although small, might be
detectable by the Planck satellite for reasonable values of the amplitude
of the present day mass fluctuations (\ie values that reproduce the
present day abundance of rich clusters of galaxies).

In this paper, we have computed the effects of gravitational lensing on
both the temperature and polarisation pattern and demonstrated that
lensing can break the geometrical degeneracy inherent in the linear CMB
power spectra. We have performed a Fisher matrix analysis to show how
gravitational lensing affects estimates of cosmological parameters. The
Fisher matrix requires derivatives of the CMB power spectra with respect
to the cosmological parameters. Since numerical errors in these
derivatives can artificially break real parameter degeneracies, we have
made a detailed analysis of numerical errors in our computations and shown
that they are small.

The results of our Fisher matrix analysis are summarized in Tables 3
\&\ 4 for an idealized two dimensional space of $\omega_\Lambda$ and
$\omega_K$ and for a more realistic space of six cosmological
parameters.  These show that gravitational lensing is detectable by a
Planck-type experiment and must be taken into account when estimating
the values of cosmological parameters. The effects of gravitational
lensing are detectable in both the temperature and polarisation
anisotropies. For some experimental parameters, the effects of lensing
are more easily detectable in the polarisation signal (because of the
sharpness of the peaks and minima in the polarisation power spectrum)
than in the temperature power spectrum, even though the anisotropies
are polarised at only the few percent level.  Gravitational lensing of
the CMB anisotropies breaks the geometrical degeneracy and so it
should be possible to set limits on the values of $\omega_\Lambda$ and
$\omega_K$ from observations of the CMB anisotropies alone. For
example, from the 6 parameter analysis in Table 4 for model 1a (a
spatially flat $\Lambda$-dominated universe) it should be possible to
set $1\sigma$ limits of $\delta \omega_\Lambda \approx 0.03$ and $\delta
\omega_K \approx 0.003$ using temperature and polarisation measurements and
limits of $\delta \omega_\Lambda \approx 0.04$ and 
$\delta \omega_K \approx 0.004$  from observations of temperature
anisotropies alone.  This shows that for certain target models a
Planck-type experiment is capable of setting tight limits on the
geometry of the Universe. Furthermore, the possibility of detecting
gravitational lensing adds to the  scientific case for measuring
CMB polarisation at high sensitivity and angular resolution.

The lensing constraints on $\omega_\Lambda$ and
$\omega_K$ are sensitive to the normalisation of the present day mass
fluctuations and the growth rate of the matter fluctuations hence we find
less stringent limits for a standard CDM model
normalised to $\sigma_8(t_0) = 0.52$ (Tables 3 and 4). Nevertheless, even in
this case, a Planck-like experiment can set $1 \sigma$ errors of 
$\omega_\Lambda \approx 0.06$ and $\omega_K = 0.01$.

The geometrical degeneracy can be broken by applying constraints
derived from more conventional astronomical techniques. For example,
accurate measurements of the Hubble constant, age of the Universe, the
luminosity distances of Type 1a supernovae, measurements of
large-scale galaxy clustering can be used, with various assumptions,
to break the geometrical degeneracy (see Efstathiou \&\ Bond
1998). However, as described in this paper, gravitational lensing
breaks the geometrical degeneracy and so one can disentangle the
values of $\Omega_K$ and $\Omega_\Lambda$ from accurate observations
of the CMB anisotropies.
Comparison of results of CMB-based experiments with those obtained with more conventional techniques
can provide consistency checks and tests of possible systematic
errors.

\vskip 0.1truein

\noindent
{\bf Acknowledgements}
GPE would like to thank PPARC for the award of a Senior Research
Fellowship. RS is supported by UK PPARC grant and acknowledges help of
Polish Scientific Committee (KBN) grant No. 2 P03D 008 13.X2. We thank
the referee Matias Zaldarriaga for many useful comments and for
encouraging us to include the temperature-polarisation power spectrum
in the Fisher matrix analysis of Section 4.

\end{document}